\documentstyle[aps,twocolumn,epsfig]{revtex}
\begin{document}
\draft
\preprint{}
\title{Black holes in extra dimensions can decay on the bulk}

\author{\bf A. K. Chaudhuri\cite{byline}}
\address{ Variable Energy Cyclotron Centre\\
1/AF,Bidhan Nagar, Calcutta - 700 064}

\maketitle

\begin{abstract}
In the extra dimensional theories, with TeV scale Plank constant,
black  holes  may  be  produced  in  the  Large  Hadron  Collider
experiments. We have argued that 
in the d-dimensional black hole,
the intrinsically 4-dimensional brane fields do not see the
same geometry at the horizon, as   in a
4-dimensional space-time. Kaluza-Klein modes
invades the brane and surroundings and the brane fields can be
considered as a thermal system at the temperature of the black hole.
From  energy  and  entropy
consideration,  we  show  that  whether  or not a six-dimensional
black hole will decay  by  emitting  Kaluza-Klein  modes  or  the
standard  model  particles, will depend on the length scale of the
extra dimensions as well as on the mass of the  black  hole.  For
higher  dimensional black holes, Kaluza-Klein modes will dominate
the decay.
\end{abstract}

Recently  several  authors  have proposed that in addition to the
usual 3+1 dimensions, our  universe  may  contain  extra  spatial
dimensions,  as  large as a millimeter \cite{arh98}. The proposal
gain popularity as the models with extra dimensions  could  solve
the  hierarchy  problem,  without  recourse  to  supersymmetry or
Technicolor. If  such  theories  are  correct,  then  it  may  be
possible  to  produce black holes in the laboratory conditions at
the  Large  Hadron  Collider  (LHC).  The  idea  is  simple.  The
effective  4-dimensional Plank's constant $M_4$ is related to the
fundamental   d-dimensional   Plank's    constant    $M_d$    as,
$M_d^{d-2}=M_4  L^{-(d-4)}$, $L$ being the compactification scale
of the extra dimensions. For millimeter  scale  $L$,  fundamental
Plank  constant $M_d$ could be as low as $\sim$ TeV, much smaller
than the LHC cm energy ($\sqrt{s}$ =14 TeV). Banks  and  Fischler
\cite{ba99} argued that for $\sqrt{s} >> M_{d}$, scattering cross
sections are dominated by the black hole formation, at the impact
parameters   less   than   the   Schwarzschild   radius.   Simple
calculations \cite{di01,gi02} indicate that  at  the  LHC,  black
hole production cross sections can be large.

Several  authors  have  considered  formation  and decay of black
holes        in        extra         dimensional         theories
\cite{ba99,di01,gi02,my86,ar98,ri02,ch02,em00,ca01}.   While  LHC
can produce black holes  in  abundant,  its  detection  may  face
problems.  Black holes will decay through Hawking radiation. Some
authors \cite{ba99,ar98} have argued that the d-dimensional black
holes  will  radiate  mainly  into   Kaluza-Klein   (KK)   modes.
Temperature ($T_{BH}$) of extra dimensional black holes are large
and all the KK modes below $T_{BH}$ will be produced. Emission of
standard model (SM) particles will be limited, due to phase space
reason.  Emparan,  Horowitz  and  Myers  \cite{em00} on the other
hand, argued that the  d-dimensional  black  holes  will  radiate
mainly   on   the   brane.  They  argued  that  the  phase  space
consideration could not be applied to  SM  fields,  as  they  are
intrinsically  4-dimensional.  The  4-dimensional  black hole law
governs their emission. For black hole radius  $r_d  <<  L$,  the
emission of KK modes will be suppressed by a factor of $(r_d/L)$,
compared  to  SM particles. They compared emission rates of black
holes in d-dimension and in 4-dimension and showed that the ratio
of emission rates gets closer to one as dimensionality increases.
Thus, $(dE_4/dt)/(dE_6/dt) \sim 3.66$ and $(dE_4/dt)/(dE_{10}/dt)
\sim 1.54$ \cite{em00}.

Whether or not d-dimensional black holes decay on the bulk or 
on the brane is an important issue as the detection of laboratory
produced black holes will depend on it. If they decay on the bulk,
they will remain undetected. 
In the d-dimensional black hole,
the intrinsically 4-dimensional brane fields do not see the
same geometry at the horizon as   in a
4-dimensional space-time. Matter fields (i.e. Kaluza-Klein modes)
invades the brane and surroundings. In the present paper, we
assume that the brane fields can be considered as a thermal
system, in equilibrium with the matter fields at the black hole
temperature.
We  then  study  the  energy  and
entropy  of  the brane fields (i.e. SM particles) vis-a-vis Kaluza-Klein modes. It is
seen that for $d>6$, energy and entropy in the Kaluza-Klein modes
greatly exceeds the energy and entropy of the SM particles. It is
then expected that $d>6$ black holes will decay mainly in to  the
bulk,  by  emitting  KK  modes.  For  $d=6$ black holes, KK modes
dominate the energy and entropy above a critical compactification
scale, which depend on the black hole mass. Below  that  critical
compactification  scale,  SM  particles  dominate  the energy and
entropy of the black holes.

Let  us assume a mass $M$ is collapsed in to a black hole of size
$r_d << L$, in $d>4$ dimension. The extra-dimensions are  of  the
scale  $L$.  The d-dimensional metric (extension of Schwarzschild
solution in d-dimension) can be written as \cite{my86},

\begin{equation}
ds^2 = -f(r) dt^2 + f^{-1}(r) dr^2 +r^2 d\Omega^2_{d-2},
\end{equation}

\noindent  with  $f(r)=1-(r_d/r)^{d-3}$.  $d\Omega^2_{d}$  is the
line element on the unit d-sphere. The horizon lies  at  $r=r_d$,
the black hole radius  \cite{my86},

\begin{equation}
r_d = \left[\frac{16 \pi G_d M}{(d-2) A_{d-2}}\right]^{\frac{1}{d-3}}
\end{equation}

\noindent  where $G_d$ is the d-dimensional Newton's constant and
$A_{d-2}$ is the area of the unit (d-2)-dimensional  sphere.

\begin{equation}
A_{d-2}=\frac{2\pi^{(d-1)/2}}{\Gamma(\frac{d-1}{2})}
\end{equation}

The  induced metric in 4-dimension \cite{em00},

\begin{equation}
ds^2 = -f(r) dt^2 + f^{-1}(r) dr^2 +r^2 d\Omega^2_{2},
\end{equation}

\noindent is of the same form as the d-dimension metric, with
$f(r)$  same  as  before. The event  horizon $r=r_d$ is certainly
not of 4-dimensional Schwarzschild geometry. Ricci tensor of this
four dimensional metric is non-zero near the horizon. It can be thought
upon as a black hole with matter fields (kaluza-Klein modes) around it.
This observation led us to assume
that the brane fields can be considered as a thermal system of SM
particles.  The  temperature of the thermal system is the same as
the  black  hole  temperature.  The  gravity   being   the   only
interaction  common  to the brane and the bulk, the brane will be
populated  by  the  KK  modes,  at   the   temperature   of   the
d-dimensional black hole. Interaction of the KK modes with the SM
particles  can  thermalise  them. As the d-dimensional gravity is
much stronger than  the  4-dimensional  gravity,  KK  modes  will
interact much more strongly than the usual gravitons. Lifetime of
d-dimensional  black holes being large, the assumption of thermal
equilibrium  between  the  KK  modes  and  the  SM  particles  is
reasonable.

If we approximate $(dE_d/dt)/dE_4/dt) \sim E_d/E_4$, then whether
the  black  hole  will  decay  in  to the brane or the bulk, will
depend on the relative energy $E_d$ and $E_4$. As will  be  shown
below, apart from the black hole mass, it depend on the number of
extra  dimension  as well as on the compactification scale of the
extra-dimensions.

Energy  density ($\varepsilon$), pressure (p) and entropy density
($\sigma$) of the SM particles at a temperature  of  $T$  can  be
easily  calculated  for  a  non-interacting  system.  For massive
particles,

\begin{eqnarray*}
\varepsilon_{mass} (T) = &&
\sum_i \frac{g_i}{(2\pi)^3} \int  \frac{\sqrt{k^2+m_i^2}}
{exp(\frac{\sqrt{k^2+m_i^2}}{T}) \pm 1} d^3k  \\
p_{mass} (T) = &&
\sum_i \frac{g_i}{(2\pi)^3} \int  \frac{k^2}{3\sqrt{k^2+m_i^2}}
\frac{1}{exp(\frac{\sqrt{k^2+m_i^2}}{T}) \pm 1}  d^3k\\
\sigma_{mass}(T) = && \frac{\varepsilon_{mass} + p_{mass}}{T}
\end{eqnarray*}

\noindent  where  $g_i$ is the degeneracy of the particle of mass
$m_i$ and the $\pm$ is for boson/fermion. The energy  density  of
mass   less   particles   is  calculated  using  the  $T^4$  law,
($\varepsilon_{mass  less}=\pi^2  N  T^4/30$),  $N$   being   the
effective  degrees  of  freedom).  We  also  assume the ideal gas
equation of state,  $p=1/3  \varepsilon$  for  them.  In  the  SM
particle  lists we have included six quarks (u,d,s,c,b and t) and
their anti particle, three electrons ($e,\mu  ,\tau$)  and  their
anti  particles  and  $W^{\pm},  Z$  bosons. The color degrees of
freedom as well as the spin degeneracy is taken into account.  In
the  mass less particles lists, we include $8\times 2$ gluons and
$2$ photons. In Fig.1, we have shown the energy density, pressure
and the entropy density of the SM particles as a function of  the
temperature.  For the SM particles listed above, for temperatures
above 50 GeV, the energy density and pressure are well  described
by  the  $T^4$  law. The entropy density is then described by the
$T^3$ law.

The  energy  and  entropy of the SM particles, in a d-dimensional
black hole of radius $r_d$ is then obtained as,

\begin{eqnarray}
E_{SM} = && [\varepsilon_{mass} +\varepsilon_{massless}]
\frac{4\pi}{3} r^3_d\\
S_{SM} = && [\sigma_{mass} +\sigma_{massless} ]\frac{4\pi}{3} r^3_d
\end{eqnarray}

Black  hole  temperatures  being  $T_d  =(d-3)/4\pi  r_d$, in six
dimension, total entropy of the SM particles  is  independent  of
the  black  hole mass or radius. The energy and entropy due to KK
modes is then obtained as,

\begin{eqnarray}
E_{KK}=&&M_{BH} - E_{SM}\\
S_{KK}=&&S_d -S_{SM}
\end{eqnarray}

\noindent  where $M_{BH}$ is the black hole mass and $S_d$ is the
entropy of the d-dimensional black hole,

\begin{equation}
S_d = \frac{r^{d-2}_d A_{d-2}}{4G_d}.
\end{equation}

In   Fig.2,   we   have   shown   the  ratio  of  the  entropies,
$S_{KK}/S_{SM}$   (panel   a)   and   the   ratio   of   energies
$E_{KK}/E_{SM}$  (panel  b) as a function of the compactification
radius $L$ for a black hole formed  in  d=6  dimension.  We  have
considered black hole masses of 1,3,5,7, and 9 TeV, accessible at
the  LHC.  The  ratios  indicate that whether or not SM particles
will   dominate   the   emission   spectrum   depend    on    the
compactification  radius  as  well as on the black hole mass. For
$M_{BH}$=1 TeV, energy available  to  KK  modes  exceeds  the  SM
energy for $L>$ 0.2 mm. Nearly similar value is obtained from the
entropy  consideration.  Entropy  due  to  KK  modes  exceeds the
entropy due to SM particles for $L>$ 0.4  mm.  For  higher  black
hole masses, the cross over from KK mode dominance to SM particle
dominance  will  occur  at  still  lower  $L$.  We  note that the
critical  compactification  radius,  above  which  the  KK  modes
dominate the SM particles, are indicative only. We only intend to
show  that  the  emission from d=6 dimensional black holes can be
dominated by the KK modes or by the SM particles. Which one  will
dominate will depend on the compactification radius as well as on
the mass (radius) of the black holes.

Situation  is  completely  different for higher dimensional black
holes. In Fig.3, we have  shown  the  results  obtained  for  d=8
dimensional   black   holes.   It   can   be  seen  that  if  the
compactification scale lies  between  $10^{-3}$mm-  $10$  mm,  KK
modes   will   dominate   the  black  hole  energy  and  entropy.
Consequently emission will be dominated by the KK modes.

To  conclude,  we  have   approximated the
brane fields as a thermal system of standard model particles,  in
contact  with  the  Kaluza-Klein  modes.  We  have calculated the
energy density, pressure and entropy density of SM particles.  It
was  seen that for six dimensional black holes, whether the decay
will be dominated by the KK modes or by the SM particles  depends
on the compactification scale of the extra dimensions, as well as
on  the  mass  of  the black holes. For 1 TeV black holes, if the
compactification scale is below 0.2-0.4  mm,  SM  particles  will
dominate  the  decay.  For  higher mass black holes, SM particles
will dominate, if the  compactification  scale  is  still  lower.
Other  wise,  a  six  dimensional  black  hole  will decay mainly
through emission of KK modes. For $d>6$ dimensional black  holes,
the decay will be dominated by the KK modes.

\begin{figure}[h]
\centerline{\psfig{figure=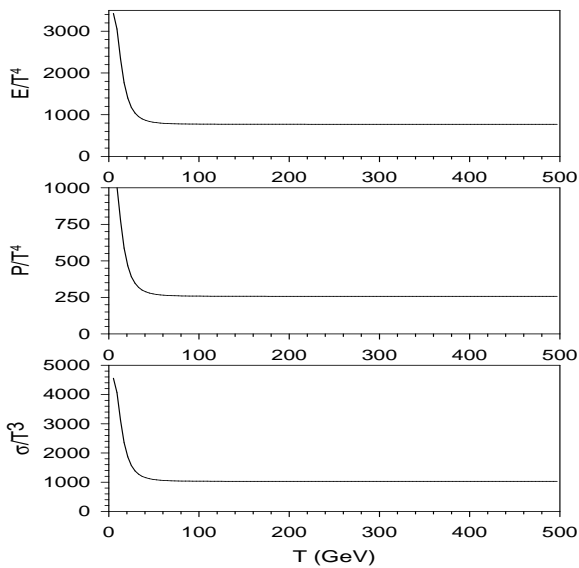,height=8cm,width=8cm}}
\caption{In three panels, energy density, pressure
and entropy density due to SM particles as a function of temperature
are shown.}
\end{figure}

\begin{figure}[h]
\centerline{\psfig{figure=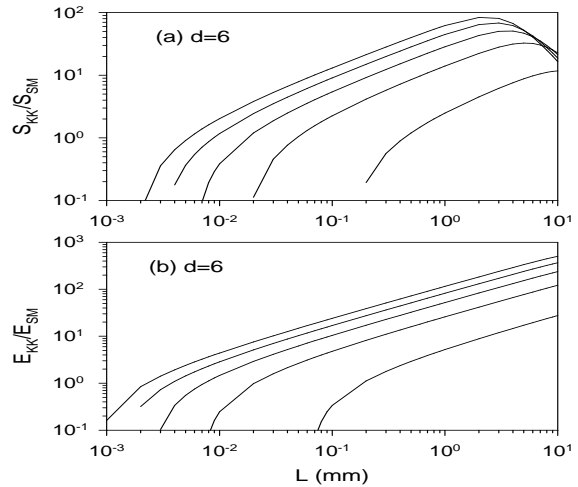,height=8cm,width=8cm}}
\caption{(a)  Ratio  of  entropy due to KK modes and SM particles
for six-dimensional black holes of mass  $M_{BH}$=1,3,5,7  and  9
TeV  (from  bottom  to  top)  are  shown,  as  a  function of the
compactification radius. (b) Ratio  of  energy  available  to  KK
modes   and   SM   particles.}
\end{figure}

\begin{figure}[h]
\centerline{\psfig{figure=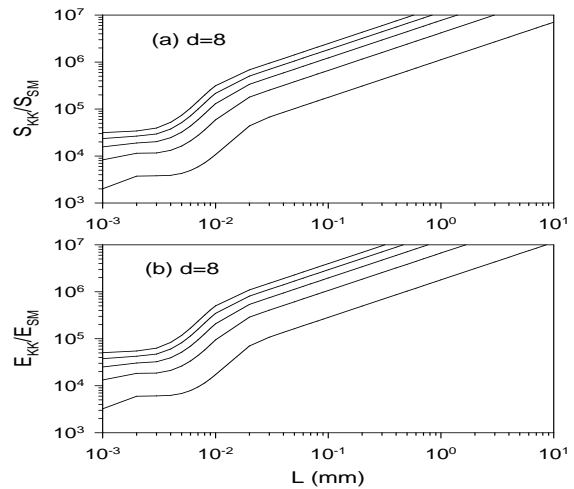,height=8cm,width=8cm}}
\caption{Same  as  Fig.2, for 8-dimensional black
holes.}
\end{figure}

\end{document}